\documentclass[aip,twocolumn,reprint,preprintnumbers,amsmath,amssymb]{revtex4-1}
\usepackage{graphicx}
\usepackage{dcolumn}
\usepackage{bm}
\usepackage{amssymb}
\usepackage{epsfig}

\begin{document}



\title{Fully-Balanced Heat Interferometer}

\author{M. J. Mart\'{\i}nez-P\'erez}
\email{mariajose.martinez@sns.it}
\affiliation{NEST, Istituto Nanoscienze-CNR and Scuola Normale Superiore, I-56127 Pisa, Italy}

\author{F. Giazotto}
\email{giazotto@sns.it}
\affiliation{NEST, Istituto Nanoscienze-CNR and Scuola Normale Superiore, I-56127 Pisa, Italy}




\begin{abstract}

A tunable and balanced heat interferometer is proposed and analyzed. The device consists of two superconductors linked together to form a double-loop interrupted by three parallel-coupled Josephson junctions. Both superconductors are held at different temperatures allowing the heat currents flowing through the structure to interfere. We demonstrate that thermal transport is coherently modulated through the application of a magnetic flux. Furthermore, such modulation can be tailored at will or even suppressed through the application of an extra control flux. Such a device allows for a versatile operation appearing as an attractive key to the onset of low-temperature coherent caloritronic circuits.

\end{abstract}


\maketitle



Manipulation of heat currents in mesoscopic circuits represents a youthful field of research.\cite{GiazottoRev, Dubi} The role of dissipative phenomena at the nanoscale and the benefits that can be derived from them are becoming to be understood. Starting from cooling applications and fine temperature tuning in cryogenic detectors,\cite{GiazottoRev} or superconducting circuits for quantum computing,\cite{NielsenChuang} and ending with the emergence of caloritronic circuits.\cite{Blickle,Zwol,Ojanen,Segal,Saira,PekolaGiazotto,PekolaPLR07,Casati,Panaitov,Ryazanov} Towards this end, a \textit{magnetic flux-controllable} superconducting heat interferometer was recently theoretically conceived\cite{GiazottoAPL12} and subsequently realized experimentally.\cite{Giazottoarxiv} This achievement served to show, on the one hand, how quantum coherence between two weak-linked superconducting condensates extends also to dissipative observables such as heat current.\cite{MakiGriffin} On the other, this device might constitute the building block for the implementation of superconducting hybrid coherent caloritronic circuits like, for instance, thermal modulators, heat transistors and splitters, etc. In this Letter we propose a step forward this goal by envisioning and theoretically analyzing a fully-balanced heat interferometer. Compared to the previous realization,\cite{GiazottoAPL12, Giazottoarxiv} our device provides an enhanced control over the flux-to-heat current transfer function therefore enabling the choice of convenient operation points for different applications. Additionally, this fully-balanced heat interferometer is much more robust against fabrication deficiencies such as differences between the normal-state resistances of the Josephson junctions. Such deficiencies can be now "corrected" during the operation therefore allowing the maximization of the phase-dependent component of heat current ($J_{int}$) or its complete annihilation.

\begin{figure}[t]
\includegraphics[width=6cm]{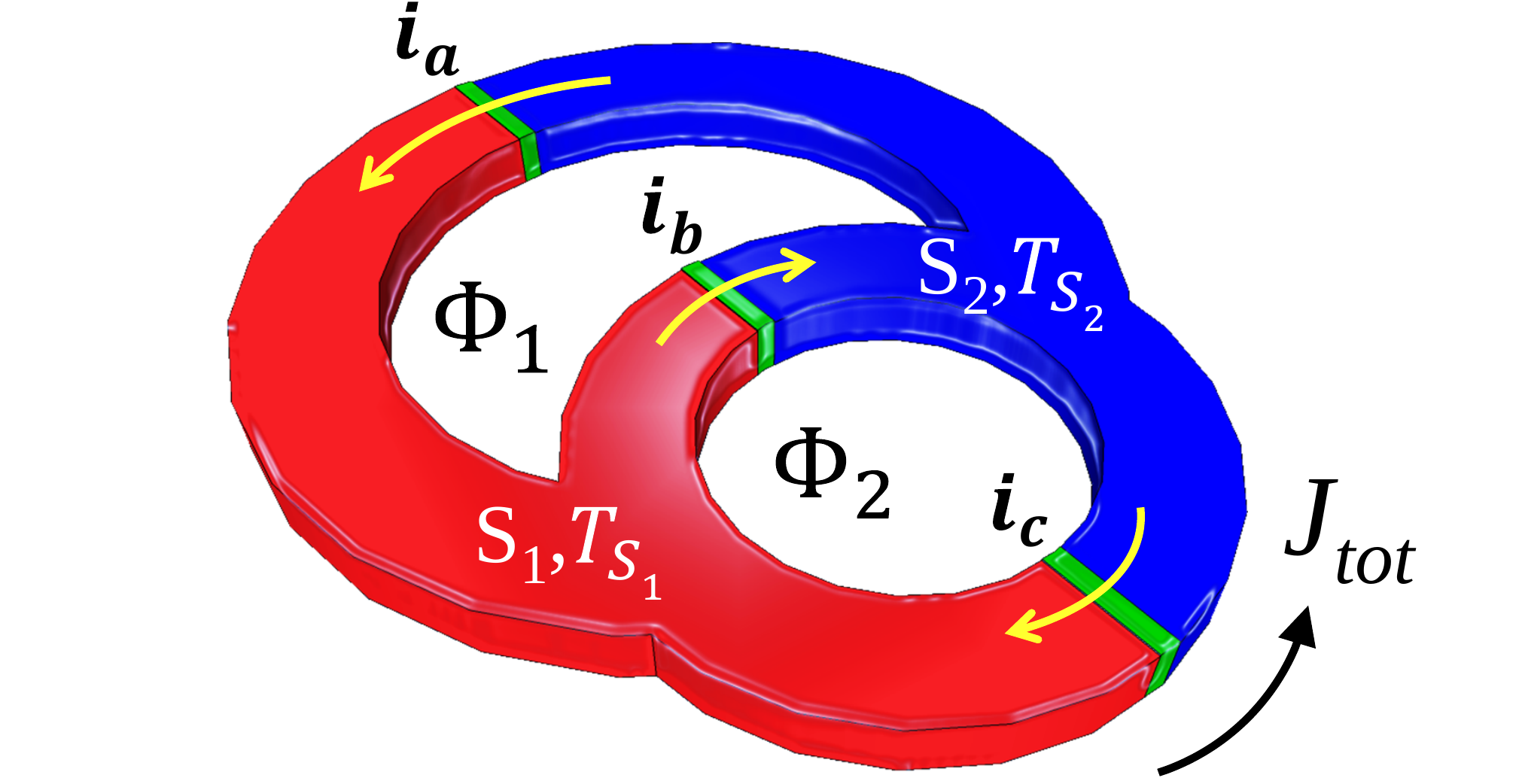}
\caption{Our superconducting circuit consists of three parallel Josephson junctions (characterized by their Josephson currents $i_a$, $i_b$ and $i_c$, respectively) that define a double-loop heat interferometer. The temperature in S$_1$ is risen up to $ T_{\texttt{S}_1}\geq T_{\texttt{S}_2}$ which yields a steady-state heat current $J_{tot}$ flowing from S$_1$ to S$_2$. Moreover, the voltage drop across the device is set to zero. The main loop is pierced by a magnetic flux $\Phi_1$ whereas a control flux $\Phi_2$ is applied through the second loop enabling enhanced performance.}
\label{Fig1}
\end{figure}

Our thermal circuit consists of two superconductors S$_1$ and S$_2$, weak linked forming a double-loop interrupted by three parallel Josephson junctions (see Fig. \ref{Fig1}). Let us denote $R_i$ the normal-state resistance of junction $i$, and $\varphi_i$ the macroscopic phase difference across junction $i$. This structure behaves as a conventional superconducting quantum interference device (SQUID) pierced by a magnetic flux $\Phi_1$ in which one of the junctions has been replaced by a DC SQUID. The characteristics of this second ``junction'' can be tuned thanks to the application of a control magnetic flux $\Phi_2$. The system is temperature biased by setting the temperature in S$_1$ to be $T_{S1}\geq T_{S2}$, $T_{S2}$ being the temperature in S$_2$. Furthermore, the voltage drop across the whole structure is set to zero. Under these circumstances, a thermal gradient arises across the junctions and a stationary heat current $J_{tot}$ will flow from S$_1$ to S$_2$, which are in steady-state thermal equilibrium.\cite{MakiGriffin,Guttman97,Zhao03} As it was argued in Ref. \citenum{GiazottoAPL12}, $J_{tot}$ results from the sum of two terms, $$ J_{tot} =  J_{qp}(T_{S_1}, T_{S_2})-J_{int}(T_{S_1}, T_{S_2},\varphi_a,\varphi_b,\varphi_c).$$ Here, $J_{qp}$ is the heat current carried by quasiparticles that depends only on the temperatures of both superconductors. On the other hand, $J_{int}$ is the interference component of the heat current that depends on the temperature of both superconductors and on the macroscopic phase difference across each junction\cite{MakiGriffin} as well. This phase-dependent term is peculiar of weakly-coupled superconductors, and arises as a consequence of the interplay between Cooper pairs and quasiparticles on tunneling events through Josephson junctions. The Cooper pair condensate carries no entropy therefore leading to a zero contribution to the heat current.\cite{MakiGriffin,Guttman97}

In the following we shall concentrate on the phase-dependent heat current only. Taking into account that we are dealing with a three-junction circuit, $J_{int}$ reads
\begin{equation}
 J_{int}(T_{S_1}, T_{S_2},\varphi_a,\varphi_b,\varphi_c) = \sum_{i=a,b,c} J_{int}^i(T_{S_1}, T_{S_2}) \cos \varphi_i.
\label{eq1}
\end{equation}
We emphasize that, whereas charge current depends on the \textit{sine} of the phase difference across a Josephson junction,\cite{Tinkham} heat current depends on the \textit{cosine} of $\varphi_i$. In Eq. (\ref{eq1}), $  J_{int}^i = \frac{ 1}{e^2 R_i }    \int _0 ^\infty d \varepsilon\varepsilon  \mathcal{M}_1(\varepsilon , T_{S_1})  \mathcal{M}_2   (\varepsilon , T_{S_2})  [ f(T_{S_2}) -  f(T_{S_1})]$ where  $ \mathcal{M}_j(\varepsilon , T_{S_j}) = \Delta_j(T_{S_j}) / \sqrt{\varepsilon^2-\Delta_j(T_{S_j})^2} \Theta [ \varepsilon^2-\Delta_j(T_{S_j})^2 ]$,  $ f(T_{S_j})=  \tanh(\varepsilon / 2 k_{\texttt{B}} T_{S_j})$, $\Delta_j(T_{S_j})$ is the temperature-dependent energy gap of superconductor $S_j$, $\Theta (x)$ is the Heaviside step function, $k_{\texttt{B}}$ is the Boltzmann constant and $e$ is the electron charge. For the sake of completeness we also provide with the expression for the heat current carried by quasiparticles,\cite{MakiGriffin}  $ J_{qp}(T_{S_1}, T_{S_2})= \sum_{i=a,b,c} J_{qp}^i(T_{S_1}, T_{S_2})$, where  $ J_{qp}^i = \frac{ 1 }{e^2 R_i }    \int_0 ^\infty d \varepsilon\varepsilon  \mathcal{N}_1(\varepsilon , T_{S_1})  \mathcal{N}_2   (\varepsilon , T_{S_2})  [ f(T_{S_2}) - f(T_{S_1}) ] $. Here  $ \mathcal{N}_j(\varepsilon , T_{S_j}) =|\varepsilon| / \sqrt{\varepsilon^2-\Delta_j(T_{S_j})^2}\Theta [ \varepsilon^2-\Delta_j(T_{S_j})^2 ] $ is the BCS normalized density of states in $S_j$.

\begin{figure}[t]
\includegraphics[width=7.6cm]{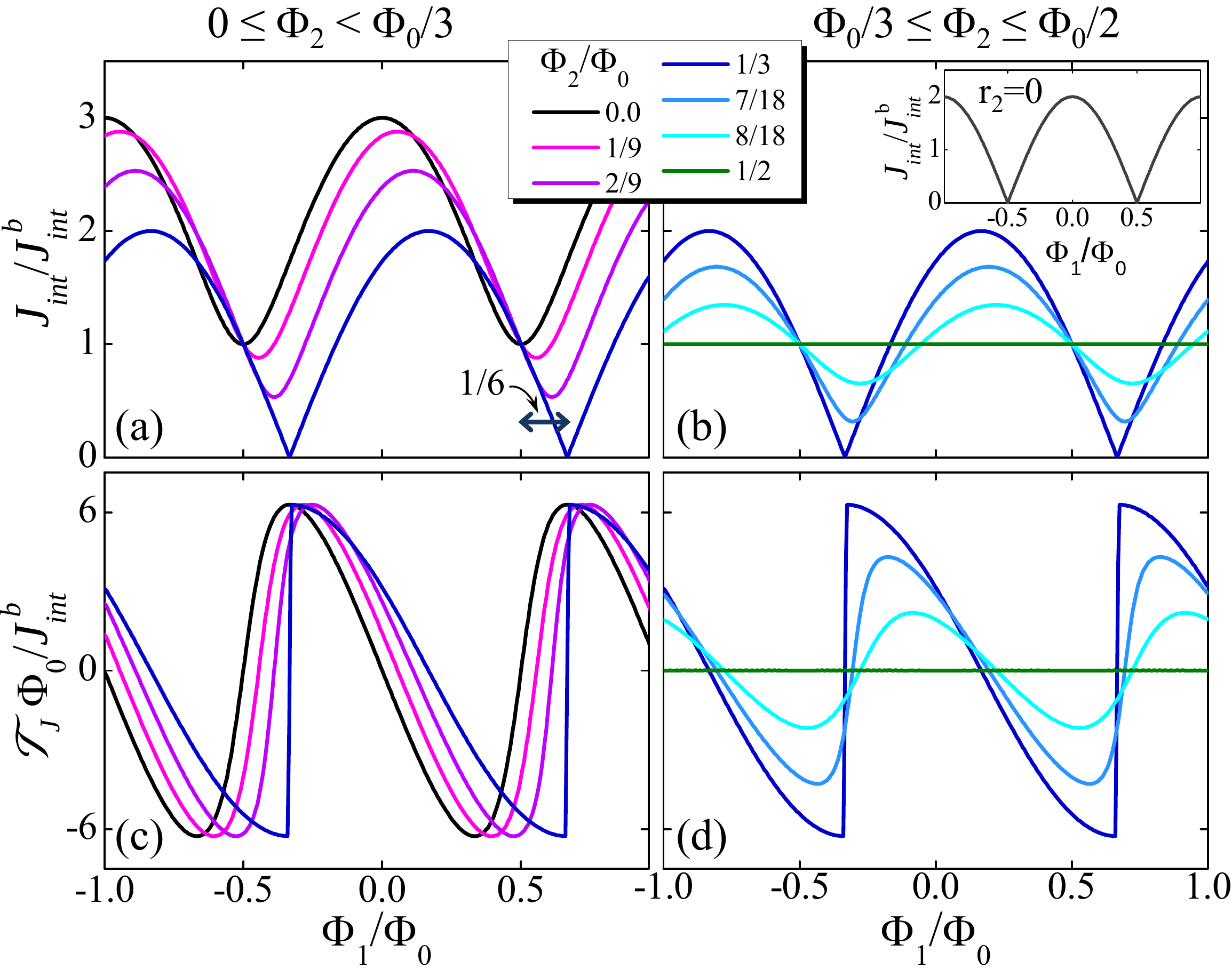}
\caption{Top panels: phase-dependent component of the heat current $J_{int}$ as a function of the magnetic flux in the main ring $\Phi_1$ plotted for different values of the control flux $\Phi_2$ for a double-loop heat interferometer with $r_1=1$ and $r_2=1$, i.e., $R_a=R_b=R_c$. The inset shows the case for $r_2=0$. Bottom panels: transfer function $\mathcal{T} _J = \partial J_{int} / \partial \Phi_1 $ vs. $\Phi_1$ plotted for the same values of $\Phi_2$ as in the top panels. As it can be seen in panels (a) and (c), the amplitude of the oscillation $\delta J_{int}$ remains constant for $0\leq \Phi_2 \leq \Phi_0/3$ decreasing for higher values of $\Phi_2$ [see panels (b) and (d)].}
\label{Fig2}
\end{figure}

The explicit dependence of $J_{int}$ on the applied magnetic fluxes and the circuit symmetry parameters, i.e., on the ratio between the normal-state resistances of the junctions, can be calculated analytically as follows. On the one hand the fluxoid quantization on both loops imposes
\begin{equation}
\begin{array}{lcll}
  \varphi_a + \varphi_b + 2 \pi \Phi_1/\Phi_0 =& 2 n \pi \\
  \varphi_b + \varphi_c + 2 \pi \Phi_2/\Phi_0 =& 2 m \pi,
\end{array}
\label{fluxquant}
\end{equation}
where $\Phi_0=2.067\times10^{-15}$ Wb is the flux quantum, and $n$ and $m$ are integers. In these expressions we have neglected the geometric inductance of the SQUID which means neglecting the self-induced flux in the loops. In a practical situation, these loops can be designed so to exhibit inductances $L_j$ of the order of a few tens of pH where $j = 1$, $2$ refers to the loop pierced by $\Phi_1$ and $\Phi_2$, respectively. Such geometry ensures a sufficiently good coupling between the SQUID loops and the additional control coils that couple $\Phi_1$ and $\Phi_2$ while providing reasonably low self-inductances.  Assuming  that each Josephson junction $i$ attains a maximal critical current $i^i_J$ of the order of   $10^2$  nA one obtains a total screening parameter $\beta = 2 L_j i^i_J/ \Phi_0 \sim 10^{-3} << 1$ allowing us to neglect the SQUID inductance. On the other hand, the consideration of finite $L_j$ makes the deduction of analytical expressions impossible and does not contribute to the understanding of the essential physics in our device. The conservation of the circulating supercurrent in both loops, on the other hand, imposes
\begin{equation}
  i^a_J \sin \varphi_a =  i^b_J \sin\varphi_b - i^c_J \sin\varphi_c,
\label{currentcons}
\end{equation}
where, according to the generalized Ambegaokar-Baratoff model,\cite{GiazottoJAP05,Tirelli} $i^i_J$ is proportional to $R^{-1}_i$. In writing Eq. (\ref{currentcons}) we have established a given current sign convention (see yellow arrows in Fig. \ref{Fig1}). 
 We define now the circuit symmetry parameters as $r_1 = i^a_J/i^b_J = J_{int}^a/J_{int}^b = R_b/R_a \geq0$, and $r_2 = i^c_J/i^b_J = J_{int}^c/J_{int}^b = R_b/R_c $. $r_2$ can vary between $0 \leq r_2 \leq 1$ since setting $r_2 >1$ is equivalent to exchange the roles of $R_b$ and $R_c$. Combining Eqs. (\ref{fluxquant}) and (\ref{currentcons}) and using simple trigonometric relations one gets
\begin{eqnarray}
 \cos \varphi_a = \pm\frac{r_1 + \alpha + r_2 \gamma}{\sqrt{1+r^2_1+r^2_2+2r_1 \alpha+2r_2 \beta +2r_1 r_2 \gamma}}  \nonumber \\
 \cos \varphi_b = \pm\frac{1 + r_1 \alpha + r_2 \beta}{\sqrt{1+r^2_1+r^2_2+2r_1 \alpha+2r_2 \beta +2r_1 r_2 \gamma}} \label{cosenos}  \\ 
 \cos \varphi_c = \pm\frac{r_2 + \beta + r_1 \gamma}{\sqrt{1+r^2_1+r^2_2+2r_1 \alpha+2r_2 \beta +2r_1 r_2 \gamma}}, \nonumber
\end{eqnarray} 
where $\alpha= \cos (2 \pi \Phi_1/\Phi_0)$, $\beta= \cos (2 \pi \Phi_2/\Phi_0)$ and $\gamma=\cos [ 2 \pi(\Phi_1 -  \Phi_2)/\Phi_0]$. The choice of the signs of Eqs. (\ref{cosenos}) obeys the requirement of minimizing the free energy ($E_J$) of the whole system. The latter can be written as $E_J = 3E_{J,0}- E_{J,0}\sum _{i=a,b,c}i^i_J \cos \varphi_i$, where $E_{J,0}=\Phi_0/ 2 \pi$.\cite{Tinkham} By inserting Eqs. (\ref{cosenos}) into the previous expression it can be easily seen that the minimum of $E_J$ is obtained for the solutions with positive signs. These, inserted into Eq. (\ref{eq1}), yield finally the following expression for $J_{int}$,
\small \begin{equation}
  J_{int}=J^{b}_{int} \sqrt{1+r^2_1+r^2_2+2r_1 \alpha+2r_2 \beta +2r_1 r_2 \gamma}. 
\label{blabli}
\end{equation}
We note that, if we set $r_2=0$, i.e., $R_c\rightarrow\infty$, one recovers the expression corresponding to the single-loop heat interferometer conceived in Ref. \citenum{GiazottoAPL12}. Consequently, for the case in which $r_1=1$ and $r_2=0$, i.e., $R_a=R_b$ and $R_c\rightarrow\infty$, Eq. (\ref{blabli}) becomes equal to $J_{int}=2 J^{b}_{int}  | \cos( \pi \Phi_1/\Phi_0) |$ [see the inset in Fig. \ref{Fig2}(b)]. This function is maximized for $\Phi_1=k \Phi_0$ and minimized for $\Phi_1=k \Phi_0/2$, $k$ being an integer. The amplitude of the oscillation $\delta J_{int}$, defined as the difference between the maximum and minimum value of $J_{int}$, is given by $2 J_{int}^b$ in this case.

\begin{figure}[t]
\includegraphics[width=8cm]{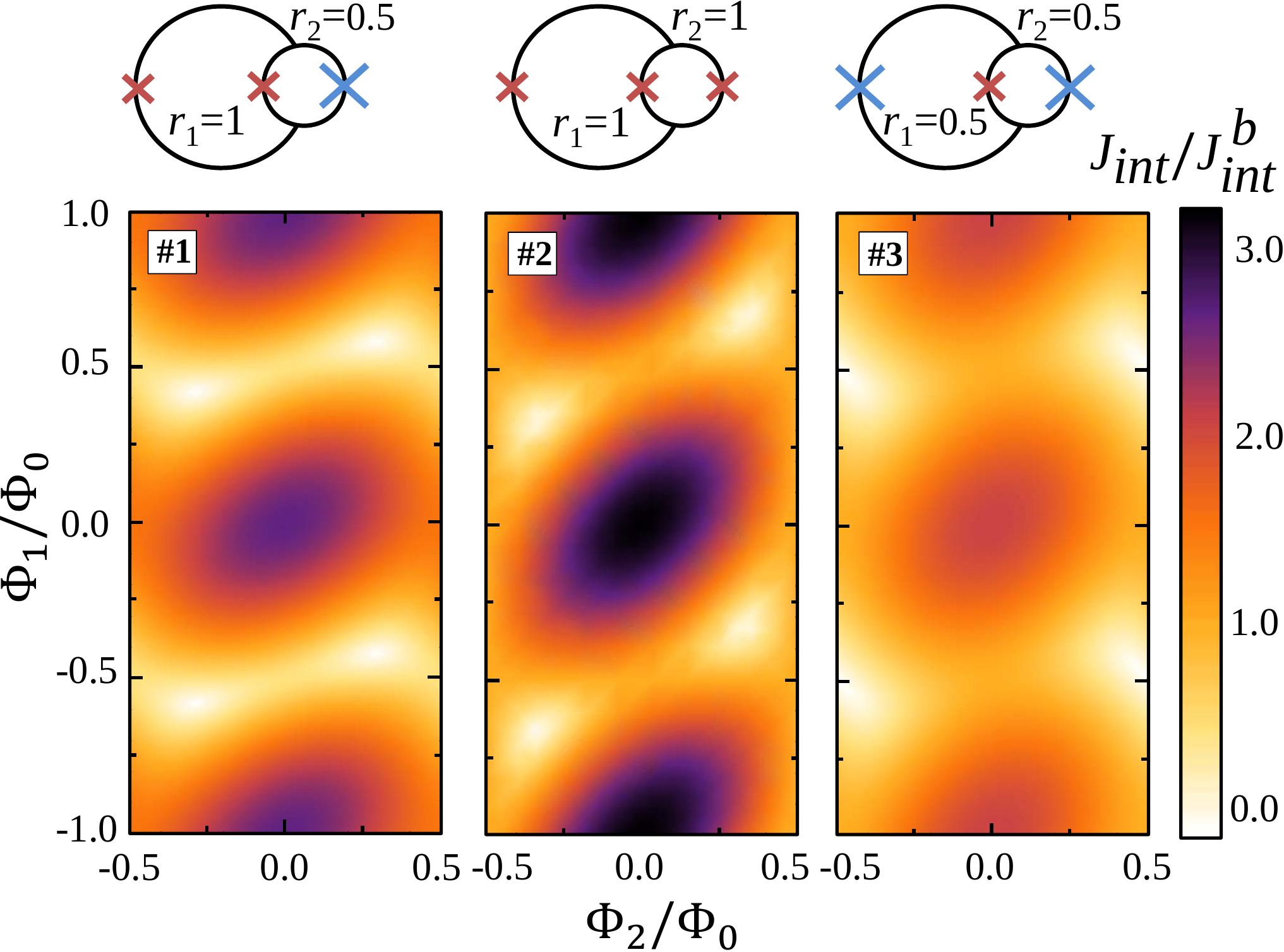}
\caption{Density plots showing $J_{int}$ as a function of the magnetic flux on the main and the control loop, $\Phi_1$ and $\Phi_2$, respectively. Three representative cases (denoted $\#1$, $\#2$ and $\#3$) have been considered and are  schematized on the top part of each plot. The maximum and minimum values of $J_{int}$ change perceptibly when reducing the symmetry of the double-loop. }
\label{Fig3}
\end{figure}

\begin{figure}[t]
\includegraphics[width=6.5 cm]{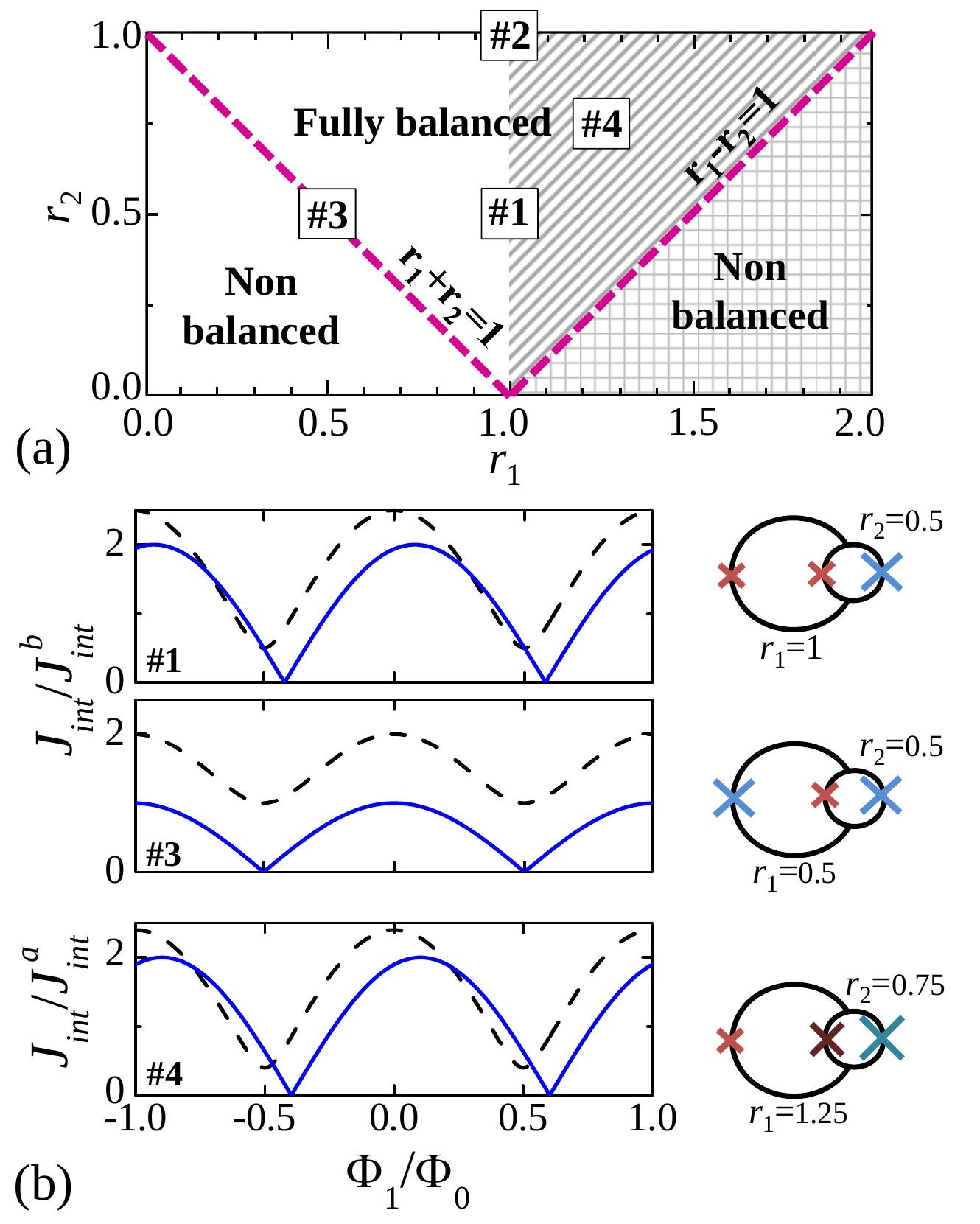}
\caption{(a) Phase diagram for the double-loop heat interferometer on the $r_1 - r_2$ space. Within the white region the amplitude of the oscillation $\delta J_{int}$ increases linearly with $r_1$. On the diagonally stripped region the maximum value of $\delta J_{int}$ is obtained independently of $r_1$ and $r_2$. Finally, on the grid region, $\delta J_{int}$ decreases with $r_2$. Additionally, the two dashed lines define a triangle within which the interferometer can be fully balanced. (b) $J_{int}$ vs. $\Phi_1$ for three selected points ($\#1$, $\#3$ and $\#4$) contained within the aforementioned triangle and sketched on the right. In the three cases, the dashed line corresponds to $\Phi_2=0$ whereas the solid line corresponds to the value of $\Phi_2$ that provides the fully suppression of $J_{int}$.}
\label{Fig4}
\end{figure}

We now turn our attention towards the double-loop heat interferometer discussed here. The straightest choice is $r_1=r_2=1$, i.e., $R_a=R_b=R_c$. Although being the most simple configuration, it enables us to infer most of the characteristics of our thermal interferometer. We can distinguish between two regimes, one defined by $0 \leq \Phi_2 < \Phi_0 /3$ [Fig. \ref{Fig2}(a)] and a second one covered by $\Phi_0/3\leq\Phi_2\leq\Phi_0 /2$ [Fig. \ref{Fig2}(b)]. If no magnetic flux is applied to the control loop, i.e., $\Phi_2=0$, the phase-dependent term of heat current is given by $J_{int}=J^{b}_{int} \sqrt{1 +8 \cos^2( \pi \Phi_1/\Phi_0)}$. As shown in Fig. \ref{Fig2}(a), this function exhibits the same expected $\Phi_0$-periodicity and amplitude of the oscillation as that of the limit case in which $r_1=1$ and $r_2=0$ [see the inset of Fig. \ref{Fig2}(b)] but the appearance of the oscillation is drastically different. In addition, notice that the minimum of $ J_{int}$ does not go to zero anymore, i.e., there is no possibility of suppressing the phase-dependent component of heat current. As we increase the amplitude of the control flux $\Phi_2$ the mean value and the shape of the curves continue to evolve whereas $\delta J_{int}$ holds unchanged. Furthermore, the curves turn out to be shifted horizontally. 

At $\Phi_2 = \Phi_0 / 3$ a noticeable phenomenon takes place. Under these circumstances Eq. (\ref{blabli}) takes the form $J_{int}=2 J^{b}_{int} | \cos[ \pi (\Phi_1/\Phi_0-1/6)] |$. This is to say, apart from a small shift equal to $\Phi_0 /6 $, one recovers the same dependence on $\Phi_1$ obtained for the symmetric single-loop heat interferometer. If we continue increasing the control flux we enter in a new regime. Under these circumstances it can be shown that $\delta J_{int}$ decreases linearly with $\Phi_2$ whereas the mean value of $ J_{int}$ remains constant. Furthermore, at $\Phi_2 = \Phi_0 /2 $, the modulation disappears completely and $J_{int}$ becomes independent of $\Phi_1$, i.e., $J_{int}=J_{int}^b$. The aforementioned characteristics are emphasized in Figs. \ref{Fig2}(c) and \ref{Fig2}(d) where we plot the transfer function, $\mathcal{T} _J = \partial J_{int} / \partial \Phi_1 $, for both regimes.

It is worthwhile now to analyze what happens by reducing the symmetry of our device. In Fig. \ref{Fig3} we show the density plots of $J_{int}$ vs. $\Phi_1$ and $\Phi_2$ for three representative cases, including the symmetric double-loop that we have analyzed previously. In general, the maximum of $J_{int}$ is always reduced for the cases in which one resistance is different from the others. Let us analyze in more detail what happens with $\delta J_{int}$. If $r_1 \leq1$ , i.e., $R_a\leq R_b$, we find that the maximum amplitude of oscillation at fixed $\Phi_2$ is given by $\delta J_{int}=  2J_{int}^b r_1 $ for whatever value of $r_2$, i.e., for whatever value of $R_c$. This is to say, $\delta J_{int}$ is independent of the degree of asymmetry of the control loop and increases linearly by increasing the symmetry on the main loop. On the other hand, if  $r_1> 1$ the dependence of $\delta J_{int}$ at fixed $\Phi_2$ is more complicated. For $r_1 - r_2 \leq 1$, $\delta J_{int}$ is independent from the degree of symmetry in both loops reaching it maximum value, that, in this case, is given by $\delta J_{int}=  2J_{int}^a$.  Finally, for $r_1 - r_2 > 1$, $\delta J_{int}$ decreases with $r_2$. These regimes are summarized in Fig. \ref{Fig4}(a).

Let us ascertain now whether it is always possible to suppress completely $ J_{int}$. The minimum of $ J_{int}$ can be easily determined from Eq. (\ref{blabli}). By requesting $ \partial J_{int} / \partial \Phi_1 = 0$ and $ \partial J_{int} / \partial \Phi_2 = 0$, and by summing the resulting equations we get that $r_1  \sin  (2 \pi \Phi_1/\Phi_0) + r_2 \sin  (2 \pi \Phi_2/\Phi_0) =0$. This equation gives us the family of values of $\Phi_1$ and $\Phi_2$ that maximize or minimize $  J_{int}$ for any given $r_1$ and $r_2$. For instance, if $\Phi_1 = \Phi_0 /2$, there are only two solutions, $\Phi_2 =0$ and $\Phi_2 = \Phi_0 /2$, that satisfy this condition and, in addition, correspond to a minimum. Inserting these solutions into Eq. (\ref{blabli}) and imposing $J_{int} = 0$ we obtain $r_1 - r_2 = 1$ and $r_1 + r_2 =1$, respectively. These equations define two straight lines in the $r_1 - r_2$ space  plotted in Fig. \ref{Fig4}. Within this area there exists at least one value of $\Phi_2$ that enables us to write $ J_{int}$ as a function of $ | \cos[ \pi (\Phi_1/\Phi_0-\theta)] | $ where $\theta$ is a shift in $\Phi_1$. The aforementioned conditions take a more eloquent form when expressed in terms of the Josephson critical currents of each junction, giving \begin{equation} 
i_a - i_c \leq i_b \leq i_a + i_c.
\label{condition}
\end{equation}  
Unlikely to a single-loop heat interferometer,\cite{GiazottoAPL12} even a quite asymmetric double-loop structure for which inequality (\ref{condition}) holds, offers the possibility of suppressing completely $ J_{int}$ through an appropriate choice of $\Phi_2$.

The three cases plotted in Fig. \ref{Fig3} satisfy the aforementioned conditions and shall therefore be useful to illustrate this behavior. In Fig. \ref{Fig4}(b) we plot $J_{int}(\Phi_1)$ for $\Phi_2=0$ (dashed lines) and for the corresponding value of $\Phi_2$ that provides the fully suppression of $J_{int}$ (solid lines). The curves corresponding to the symmetric double-loop have already been plotted in Fig. \ref{Fig2}(a) and \ref{Fig2}(b), and are therefore not shown. Notably, when setting $r_1=0.5$ and $r_2=0.5$, i.e., $R_a=R_c=2R_b$, $J_{int}$ cancels precisely when $\Phi_2= \Phi_0/2$ at $\Phi_1= \Phi_0/2 \pm k \Phi_0$ since this case satisfies exactly the condition $r_1 + r_2 = 1$. Notice that the total amplitude of the oscillation is reduced by one half with respect to the previous example. Let us finally consider a last illustrative case belonging to the region defined by $1 \leq r_1 \leq r_2+ 1$, i.e, the diagonally stripped region in Fig. \ref{Fig4}(a). If $r_1=1.25$ and $r_2=0.75$, i.e., $R_a=0.6R_c$ and $2R_b=0.75R_c$, although corresponding to a substantially asymmetric interferometer, it should  be possible to suppress completely $J_{int}$ while conserving the maximum amplitude of oscillation. As we can see in the bottom panel of Fig. \ref{Fig4}(b) this is exactly the case. Setting $\Phi_2= \Phi_0/4$, the phase-dependent component of heat transport cancels at $\Phi_1=3 \Phi_0 /5 \pm k \Phi_0$ with $\delta J_{int} = 2J_{int}^a$.

We shall finally dedicate a few words to some potential applications and practical aspects related to the fabrication of the heat interferometer proposed here. Our structure can be integrated within, not only superconducting elements, but also hybrid mesosocopic circuits composed of, e.g., normal metals, two dimensional electron gases and semiconductor nanowires as well. A precise control of the amount of heat flowing through such circuits is of crucial importance.\cite{GiazottoRev,Dubi} Temperature determines, for instance, the phase transition in superconductors,\cite{Tinkham} the amount of heat exchanged between electron and lattice phonons,\cite{GiazottoRev} the energy level occupation in quantum systems,\cite{NielsenChuang} or the critical current flowing through Josephson junctions.\cite{GiazottoRev} Mastering the heat current in superconducting circuits would enable the \textit{in-situ} fine tuning of radiation detectors.\cite{GiazottoRev,GiazottoRadition} Controlling the temperature of a two-level quantum system can eventually have influence on its decoherence time or contribute to its initialization in quantum computing architectures.\cite{NielsenChuang} Furthermore, fully tunable Josephson junctions of different kinds can be envisioned. In such devices, the direct relation between the electronic temperature and the critical current can be exploited to modulate the latter via the application of a magnetic flux. Unlike the usual \textit{voltage-controlled} hot-electron Josephson transistors,\cite{Savin,GiazottoHeikkila,Morpurgo} the principle of operation proposed here can lead to \textit{magnetic flux-controlled} thermal Josephson transistors. Last but not least, our device is furnished with two external control knobs that correspond to the two separately coupled magnetic fluxes. This opens the way to perform closed cycles in its control space parameters, which can eventually lead to the realization of a heat pump.\cite{ren10}  

Regarding the experimental realization of our double-loop heat interferometer we refer to the successful fabrication of analogous Josephson devices composed of two or more loops with independent magnetic flux controls operating as charge interferometers,\cite{Kemppinen, Chiarello} that prove the feasibility of this structure. On the other hand, a single-loop double-junction superconducting heat interferometer has indeed recently been realized experimentally.\cite{Giazottoarxiv} Our double-loop scheme provides further advantages whereas it does not imply extra difficulties from the point of view of the fabrication. Such structure can be easily fabricated by standard electron-beam lithography and shadow mask evaporation of superconducting metals, e.g., aluminum. Aluminum oxide for the Josephson barriers and copper for the normal metal electrodes can be used. A temperature gradient can easily be established across the superconducting double loop by intentionally heating one SQUID branch while maintaining the other well thermalized at the minimum bath temperature. Temperature detection and manipulation can be performed through two normal metal leads tunnel-coupled to one superconducting electrode of the SQUID that allow for the implementation of normal metal-insulator-superconductor thermometers and heaters.\cite{GiazottoRev}

To conclude, we have provided with all the informations required for designing a fully-balanced heat interferometer. Such a device allows, on the one hand, to modify the form and phase shift of the phase-dependent heat current by tuning the control flux. This would enable the user to choose a convenient point of operation within the available flux-to-heat current transfer characteristic. On the other hand, we have demonstrated that if condition (\ref{condition}) holds, a fully-balanced interferometer is obtained, meaning that the phase-dependent part of the heat current can be completely annihilated. 

We acknowledge C. Altimiras, M. Cuoco, T. T. Heikkil\"{a} and P. Solinas for comments, and the FP7 program No. 228464 ``MICROKELVIN'' for partial financial support.



\end{document}